\documentclass[11pt]{article}
\newif\ifPDF
\ifx\pdfoutput\undefined\PDFfalse
\else\ifnum\pdfoutput >0\PDFtrue
     \else\PDFfalse
     \fi
\fi

\ifPDF
   \usepackage{amssymb}
   \usepackage{amsfonts}
   \usepackage[pdftex]{graphicx,color}
\else
   \usepackage{amssymb}
   \usepackage{amsfonts}
   \usepackage[dvips]{graphicx}
\fi
\begin{document}
\begin{titlepage}

August 2005
\vskip 1.6in
\begin{center}
{\Large {\bf Salerno's model of DNA reanalysed: could solitons
have biological significance?}}
\\[5pt]
\end{center}

\normalsize
\vskip .4in

\begin{center}
J. D. Bashford \\
{\it School of Mathematics and Physics, University of Tasmania} \\
{\it Private Bag 37, Hobart 7001, Tasmania Australia} \\
%

\end{center}
\par \vskip .3in

\begin{center}
{\Large {\bf Abstract}}\\
\end{center}
We investigate the sequence-dependent behaviour of localised excitations 
in a toy, nonlinear model of DNA base-pair opening originally proposed by 
Salerno. Specifically we ask whether ``breather'' solitons could play a role 
in the facilitated location of promoters by RNA polymerase. 
In an effective potential formalism, we find excellent correlation between 
potential minima and {\em Escherichia coli} promoter recognition sites in the 
T7 bacteriophage genome. Evidence for a similar relationship between phage 
promoters and downstream coding regions is found and alternative reasons
for links between AT richness and transcriptionally-significant sites are 
discussed. 
Consideration of the soliton energy of translocation provides a novel 
dynamical picture of sliding: steep potential gradients correspond to 
deterministic motion, while ``flat'' regions, corresponding to homogeneous AT 
or GC content, are governed by random, thermal motion.
Finally we demonstrate an interesting equivalence between planar, breather
solitons and the helical motion of a sliding protein ``particle'' about a 
bent DNA axis.

\vspace{3cm}
\end{titlepage}

\section{Introduction}
Protein-DNA interactions play many of the fundamental roles in gene 
regulation. An understanding of the mechanisms involved in these 
processes is one of the major current goals for numerous biological sciences.
With large repositories of genetic information available - and costs
associated with difficult, highly specific experiments - the question of
how well such molecular interactions can be simulated is clearly important 
to investigate. Enzymes and many transcriptional factors are proteins, often 
composed of tens to hundreds of amino acids, while the DNA domains to which 
they bind can contain, in the case of prokaryotes, up to $10^{5}$ nucleotide bases. 
All-atom modelling of such a molecular complex, even neglecting the key roles of hydration and 
ions, is beyond current computational ability. 

An alternative, logical first step is to consider one specific kind of 
interaction, focussing exclusively on its salient features and develop
an accordingly simplified model. In this spirit, simple dynamical models of 
DNA have been studied for almost two decades, most successfully in regard
to describing denaturation experiments (see Ref. \cite{pey} for a review). The process by which the motions of small DNA molecules containing $\sim 10^{6}$ 
atoms are ``coarsened'' to the nucleotide base-pair, level has 
previously been qualitatively argued \cite{Gaeta1}. Typically the DNA 
molecule is modelled by one degree of freedom per base-pair: a radial 
``stretching''\cite{PB1} or a pendulum-like base ``flipping'' (Ref. \cite{Yak} being a recent review).

We note, in this context, that {\em ab initio} calculations of small DNA 
oligomers \cite{dinuc} suggest that base-pair motions can be accurately
approximated at the dinucleotide level in terms of two or three quasi-rigid, 
internal degrees of freedom, lending some weight to the coarse-graining 
assumptions previously made.
The motivation for these simple models is that, if experimental results 
can be described with a small number of degrees of freedom, then these 
degrees of freedom must be the dominant ones for the process in question. 
The applications of such a model are necessarily restricted to extremely 
specific instances of DNA behaviour \cite{Gaeta1}. Given that many regulatory 
processes are governed by highly-specific, localised denaturation of the 
DNA helix, it is logical to investigate sequence-dependent dynamical 
behaviours in such a setting.

In 1991 Salerno \cite{S1} proposed a base-flipping model of the bacterial
promoter DNA sequence $A_{1}$ in the T7 genome, suggesting the sequence had
special, ``dynamically active'' qualities with regard to propagating kink 
solitons. Subsequent investigations of other host-specific promoter T7 sequences \cite{S2}-\cite{S3b} made similar findings. Moreover in Ref. \cite{S3b} it was suggested that solitons could be created as conformational changes to the DNA helixdue to DNA-RNAP interactions. 
Recently the propagation of kinks through the entire T7 genome sequence has been studied \cite{kl}, although the (significant) differences between host- and phage-specific promoter sequences were neglected. Another paper \cite{sanchez} investigated, whether kinks might propagate differently in coding and non-coding sequences.  

Solitons had been previously suggested to have a role \cite{englander} in DNA
transcription in the 1980's, however at least one picture which developed 
\cite{riv}, of an RNA polymerase (RNAP) molecule ``surfing'' a 
thermally-driven region of open base pairs is inconsistent with the known 
conformational changes of RNAP and DNA which occur during open complex 
formation. Secondly, no absorption resonances have been observed in 
microwave spectroscopy of DNA \cite{edwards}, \cite{bigio}. Crucially the 
original motivation for invoking solitons: anomalously long lifetimes of 
DNA base-pair openings, was shown to result from misinterpretation of data 
\cite{gueron}. DNA solitons were vigorously dispelled by some researchers 
\cite{Kam} with the result that they remain little more than a curiosity 
outside of the nonlinear physics community.

More generally, a variety of studies \cite{ben3}, \cite{PB2}, \cite{bmc}
have suggested connections between the base-sequence dependence of helical 
thermal stability and transcriptional regulation sites.
The common feature in all these studies, soliton models included, is that 
AT-rich regions are distinguished by way of reduced thermal stability
relative to GC-rich regions. 
Of course regions rich in AT might stand out from a regulatory 
perspective for geometric, mechanical or chemical reasons. For example, tracts of A or T nucleotides carry intrinsic
curvature \cite{trif} and confer rigidity to the DNA superstructure 
\cite{anselmi} - \cite{lankas}, leading to significant departures 
from the average B-DNA form.
Finally the role of counterions in determining DNA structure such as bending
and groove width \cite{naplus} and affinity for such tracts is not yet fully 
understood. Sequence-dependent variations in the electrostatic surface of
DNA may also present a unique ``signature'' in promoter regions \cite{proz}.

It should be emphasised that the solitons we consider have nothing to do with 
thermally-driven, transient base-pair openings, the source of the original 
controversy. Protein-DNA interactions involve conformational changes of the 
DNA helix and in our opinion it is logical to investigate, if a small change
might be modelled by a structural perturbation to a regular B-DNA helix,
whether its translocation might be approximated by a soliton propagating 
through a nonlinear medium.

The structure of the paper is as follows: We briefly discuss aspects of 
protein-sliding and review the mechanism of lytic infection of {\em Escherichia coli} by T7 bacteriophage.
Assuming DNA molecules can actually support nonlinear, quasi-solitonic 
excitations like those of the simple ``base-flipping''model \cite{S1}, we discuss the kind of biological roles they might serve.
To this end we introduce the inhomogeneous Frenkel-Kontorova (IFK) model 
\cite{kbk}, the basis of Salerno's approach \cite{S1}-\cite{S3},  and its 
breather soliton solutions. 
The propagation of breather solutions is analysed via an effective potential 
formalism and  we compute the energy landscape, comparing extrema with 
bacterial and phage promoters of the T7 genome. 
Supposing then, that solitonic excitations of DNA do not exist, we discuss 
alternative reasons why the correlations obtained between regulatory features 
and potential extrema might have been obtained. In particular a novel 
equivalence between base-flipping stability in the planar IFK model and 
bending in a 3-dimensional, helical model is outlined.

\subsection{Protein sliding}
The mechanisms by which regulatory proteins, such as RNAP, can 
recognise their specific binding sites among tens, or even hundreds of 
thousands, of structurally identical nonspecific sites on a DNA strand are 
generally not well understood.  A widely accepted 
hypothesis is that many proteins have several modes \cite{vonhip} with 
which to bind to DNA:
 
Nonspecific binding occurs when a polar domain of the protein displaces 
cations in the major/minor grooves of the DNA helix. The protein effectively
slides, one-dimensionally, along the groove through a series of nonspecific 
binding events. The translocation mechanism for the nonspecific complex
is not known, but is ATP-independent \cite{proz} and widely assumed to be 
driven by thermal motion.  
Other possible diffusion modes include ``hopping'', where a nonspecifically 
bound protein dissociates and re-associates within the same DNA domain.
For some proteins, such as {\em lac} repressor \cite{lac1}, a possibility for 
transfer between sequentially-distant regions of DNA brought 
close together in 3-dimensional space also exists. In general, the 
facilitated location of an operator by a protein is likely to consist of a 
sequence of sliding, hopping or intersegmental transfer events \cite{vonhip}. 
A growing body of evidence exists that many 
regulatory proteins, such as repressors \cite{lac1}, \cite{p53} RNAP's\cite{wu}-\cite{bustamente}, nucleases \cite{ecorv} and methylases \cite{meth} locate 
their operator sites in this way.

While the sliding component of facilitated target location is commonly assumed 
to be thermally driven we point out there is no experimental evidence to 
preclude dynamical effects resulting from local, sequence-dependent mechanical
properties of DNA.
In the seminal paper of Berg and co-workers \cite{berg} there are two 
assumptions, in particular, which may be unsatisfactory. Firstly the 
facilitated transport model is derived under the assumption of 
a homogeneous, free protein distribution. While such conditions can be 
arranged {\em in vitro} this is not the case for a biological system.
Secondly there is no real account of degrees of molecular recognition: 
operator sites are treated as ``sinks'', defining a boundary condition for the 
one-dimensional diffusion equation. It is highly probable that some kind of 
``reading'' process also occurs, mediated by the electrostatic interactions 
between protein residues and nucleotide functional groups lying in one, or 
both, of the grooves.  

The limited empirical studies of sliding RNAP performed to date, for example 
Refs \cite{wu}-\cite{bustamente}, invariably average behaviour over many 
individual sliding events, masking any sequence-dependent variation which 
might exist. On the other hand a recent model of the hypothetical reading 
process \cite{S4} was built, based upon the assertion that 
\begin{quote}...the protein 
should follow a noise-influenced, sequence-dependent motion that includes the 
possibility of slowing down, pauses and stops...
\end{quote}

Let us qualitatively envisage then, how soliton-like deformations might arise 
in RNAP-DNA interactions. The presence of enzymes as ``mass defects'' in a 
1-dimensional DNA model have previously been considered \cite{satar}, \cite{ting0}, \cite{ting1} with regard to thermal breathers and transcribing RNAP.
In distinction, initial binding to a nonspecific DNA site entails 
insertion of polymerase domains into the major groove of the helix, where the 
displacement of counterions occurs. Suppose that the initial contact 
and recoil of RNAP during association induces a localised deformation in the 
B-DNA helix, which we approximate by a breather soliton.
Breather excitations in the IFK model \cite{kbk} are, due to the discreteness
of the model, inherently unstable. Therefore the initially stationary breather would 
propagate along the strand, preferentially in a direction determined by local 
inhomogeneities in the base sequence. Further, the deformation is not a true 
soliton, owing to the discreteness of the DNA lattice, and radiates 
energy, eventually dissipating. In this regard we also note the mean sliding 
distance of RNAP's are known to be highly sensitive to variations in cation
concentration \cite{wu2},\cite{smeekins}.

There are two pictures which are plausibly consistent with noisy, 
deterministic dynamics: either the RNAP can effectively ``surf'' the breather
or that randomly moving RNAP and deterministically travelling breathers can 
interact somehow on collision. Regarding the latter, little is known about the
structure of nonspecific protein-DNA complexes and for the remainder of
the paper we consider the former, less speculative, scenario.
 
\subsection{T7 bacteriophage}
Bacteriophage T7 is a member of the {\em Podovirales} family of viruses, which cause lytic infection of bacteria. Its simple regulatory apparatus is one of the most widely studied, serving as a model for genomes of more complex organisms. 
As mentioned above, previous nonlinear DNA studies \cite{S1}-\cite{S3}, \cite{kl} have 
involved the T7 phage genome sequence.
However these studies focussed exclusively on the base sequence, with no 
consideration for the possible changing biological context of the information
it contains. In fact the essentially linear processes of T7 DNA
translocation and gene expression make this phage an excellent case study. 

T7 is known to inject its double-stranded, linear DNA into a host {\em E. coli} cell in a stepwise, transcription-dependent manner \cite{inf1}. 
The T7 genome contains 39,937 base pairs but initially only the first 850 of 
these base are translocated from the phage particle \cite{inf2}. 
This initial fragment contains three strong promoters specific to {\em E. coli} RNAP $A_{1}-A_{3}$ (in addition to the minor $A_{0}$, or $D$, promoter with 
no known {\em in vivo} function) which initiate transcription of the phage 
sequence. The remainder of the genome is divided into three sections:
The ``early'' region contains class I genes - those responsible for modifying 
host metabolism to favour phage production; the middle region, where
class II genes govern phage replication; the ``late region'' of class III 
genes driving maturation and packaging of newly assembled phage DNA strands.

Transcription of the initial fragment serves a dual function: ``pulling'' 
downstream, early DNA from the phage particle into the host cell, in 
addition to transcribing the class I genes. 
The product of the first of these, gene 0.3, inactivates host defence 
(specifically type I restriction/modification) systems,  
therefore rapid recognition of a major promoter is vital for successful 
infection by wild-type T7.
Another product of this early region is a T7 RNA polymerase, recognising
its own specific promoters,  which is responsible for transcribing the 
remainder of the genome, a process which proceeds in two steps:
Entry of the middle region into the host cell is dependent upon the 
successful translation of class I genes. In turn, translocation of the 
late-transcribed region requires the products of early and mid-regions.

In contrast to gene expression in more complex organisms, there are very 
few ``feedback'' loops, indeed a virtual simulation of the T7 life cycle has 
been developed \cite{endy}. There are two known loops which may 
have relevance to our analysis below: mid-late inhibition of class I (host-specific) promoters \cite{hes} and late inhibition of the mid (phage-specific) promoters\cite{lys}.
\section{The model}
At physiological temperatures the physically dominant mode of base-pair opening is the base-flipping, pendular oscillations of bases about their N-glycosydic 
bond in the mean base-pair plane. Such models previously considered 
for biological roles \cite{S1}-\cite{sanchez} are based upon the IFK \cite{kbk} Hamiltonian:
\begin{eqnarray}
{\cal H} & = & \frac{1}{2} \sum_{i=1}^{n}
I_{i}(\dot{\theta}_{i}^{2} + \dot{\psi}_{i}^{2})
+ \frac{1}{2} \sum_{i=1}^{n-1}( \kappa_{i}
(\theta_{i+1}-\theta_{i})^{2} + \bar{\kappa}_{i}
(\psi_{i+1}-\psi_{i})^{2} ) \nonumber \\
& & + \sum_{i=1}^{n} \sigma_{i}(1-\cos(\theta_{i}-\psi_{i})). \label{ifk} 
\end{eqnarray}
Here $\theta_{i}$, $\psi_{i}$ are the angles of deflection of the
$i^{th}$ base ``pendulum'' and that of its complement from equilibrium, while
$I_{i}$ is the inertial moment. Nearest-neighbour bases are coupled by an 
harmonic torsion potential with ``stiffnesses'' $\kappa_{i}$, 
$\bar{\kappa}_{i}$. Finally $\sigma_{i}$ is the characteristic strength of the 
nonlinear H-bonding potential between complementary bases.

In earlier studies \cite{S1}-\cite{S3} homogeneous inertial moments and 
stiffness constants were assumed, with the only sequence-dependence residing 
in the H-bonding coupling constants $\sigma_{i}$. 
Specifically, for $i, 1 \ldots, n$
\begin{eqnarray*}
I_{i}=I, & &\kappa_{i} \equiv K =\bar{\kappa}_{i}.
\end{eqnarray*}
In addition it was assumed that $\sigma_{i} = \lambda_{i} k$ where $k$ is a 
generic coupling and $\lambda_{i}$ takes the values 2 and 3 for A.T and 
G.C pairs respectively, accounting for the differing numbers of base pairs.
With these approximations, one passes to angle sum- and difference-coordinates \begin{eqnarray*}
\theta_{i}=\frac{1}{2}(u_{i}+v_{i}), & & \phi_{i}=\frac{1}{2}(u_{i}-v_{i}).
\end{eqnarray*}
The Hamiltonian thus obtained is
\begin{eqnarray}
{\cal H}' & = & \frac{1}{2} \sum_{i=1}^{n}
I(\dot{u}_{i}^{2} + \dot{v}_{i}^{2})
+ \frac{K}{2} \sum_{i=1}^{n-1} (u_{i+1}-u_{i})^{2} + (v_{i+1}+v_{i})^{2} ) \nonumber \\
& & + \sum_{i=1}^{n} \lambda_{i}k(1-\cos(u_{i})). \label{ifk2} 
\end{eqnarray}
The equations of motion for the  $u_{i}$ reduce to the set of dimensionless,
coupled equations:
\begin{equation}
\ddot{u}_{i}-(u_{i+1}-2 u_{i}+u_{i-1})+\beta_{i} \sin u_{i}=0; \hspace{0.1cm}
u_{i}=\theta_{i}-\psi_{i}, \label{dsc}
\end{equation}
where the time variable has been rescaled, $t \to \sqrt{I/k}t$, and the 
parameter $\beta_{i}=\lambda_{i}k/K\equiv \lambda_{i}\eta$.
To model the sequence variation as small perturbations to a homogenous solution, we first require the average value of the parameters $\beta_{i}$:
\begin{equation}
\beta = \left(2 \frac{n_{AT}}{n} + 3  \left( 1 -\frac{n_{AT}}{n} \right)\right)\eta,
\end{equation}
where there are $n_{AT}$ occurrences of A.T pairs in the molecule. 
In a purely homogeneous approximation, $\beta_{i}\to \beta$, in the continuum 
limit the system of equations (\ref{dsc}) reduces to the sine-Gordon equation,
\begin{equation}
\ddot{u}-u'' +\beta \sin u =0, \label{sg}
\end{equation}
which has a rich variety of solitonic solutions. A family of 
``breather'' solutions of Eq.(\ref{sg}) with lengths $L_{\mu}$ and internal 
frequencies $\omega_{\mu}$ is defined by 
\begin{equation}
u_{br}(x,t)=4 \tan^{-1} \left(\frac{\sin \omega_{\mu} t}{\omega_{\mu} L_{\mu}}
\textrm{sech}(\frac{x}{L_{\mu}})\right),
\end{equation}
where in terms of the classifying parameter $\mu$
\begin{eqnarray*}
L_{\mu}=\beta^{-1/2} \textrm{cosec} \mu, & & \omega_{\mu}=\beta^{1/2} \cos\mu.
\end{eqnarray*}
Note that the above relation imposes a minimum breather width and frequency which 
the model can support for a given set of environmental conditions. 
The smallness of $\beta$ ensures that an approximate solution of the 
discrete, inhomogeneous model is of a similar form with slowly-varying
parameters, thus our {\em ansatz} for Eq.(\ref{dsc}) is
\begin{equation}
u_{n}=4 \tan^{-1} \frac{\sin \omega t}{\omega L}
\textrm{sech}z_{n}; \hspace{0.3cm} z_{n}=(n-X)/L.  \label{sech}
\end{equation}
Here $X$ is understood as a collective coordinate for the breather and, for 
convenience we have omitted the $\mu$ subscript.

If the total energy is approximately conserved, upon substituting 
(\ref{sech}) into the Hamiltonian (\ref{ifk2}) one arrives at an expression 
for the effective potential in the collective coordinate $X$ \cite{S3}, 
associated with the propagation of the initial excitation on an inhomogenous 
background. We find, using the identity
\begin{eqnarray*}
1-\cos u = 8/(\tan u/4 + \cot u/4)^{2},
\end{eqnarray*}
the expression for the total energy takes the form 
\begin{equation}
E(X;t)\equiv K(X;t)+V(X;t)+O(\beta^{2})=0.
\end{equation}
Here  
\begin{eqnarray*}
K(X;t)&=&
\frac{8}{L^{2}} \sum_{i=1}^n D_{i}(X;t) \left( \alpha(t)\sinh^{2} z_{i}(\dot{X})^{2} \nonumber \right. \\
& & \mbox{} \left. +\omega L\sqrt{\alpha(t)(\frac{1}{(\omega L)^{2}}-\alpha(t))} \sinh 2z_{i} \dot X\right), \\
V(X;t)& =& 8 \sum_{i=1}^n D_{i}(X;t) \left((\frac{1}{L^{2}}-\omega^{2}\alpha(t))\cosh^{2}z_{i} \right. \nonumber \\
& & \mbox{}\left.+ \alpha(t)(\frac{1}{L^{2}}\sinh^{2} z_{i} + \beta_{i}\cosh^{2}z_{i})\right), \\
D_{i}(X;t)& =&\frac{1}{(\alpha(t)+\cosh^{2}z_{i})^{2}}.
\end{eqnarray*}
where the function  $\alpha(t)\equiv (\sin (\omega t)/\omega L)^{2}$ governs 
the time-dependence of $V$. If the breather oscillation timescale is typically
orders of magnitude smaller than that of its propagation along the DNA 
\cite{ting1} we can replace the time-dependent potential by its average value
\cite{zhang}:
\begin{eqnarray}
V_{av}(X)& \equiv &\frac{1}{T}\int_{0}^{T} V(X; t) \nonumber \\
& = & \frac{4}{L^{2}} \sum^{n}_{i=1} 
\frac{(\textrm{sec}^{2}\mu+\beta_{i}L^{2}\tan^{2}\mu)\cosh z_{i}}{(\tan^{2}\mu+\cosh^{2}z_{i})^{3/2}} \label{vpot}
\end{eqnarray}
Owing to the nonlinear nature of the model the energy to translocate the
initial deformation is several orders of magnitude less that required to 
create the breather initially. We can derive a simple estimate of the 
``noisiness'' of the sliding dynamics from the energy required to shift the 
breather by one base-pair:
\begin{equation}
\varepsilon(X)=\frac{K}{k_{B}T} (V_{av}(X_{i+1})-V_{av}(X_{i})) \label{theta}
\end{equation}
where $V_{av}$ is the time-averaged potential (\ref{vpot}) and $k_{B}$ is 
Boltzmann's constant. For steep gradients the picture of sliding RNAP is 
thus analogous to a particle moving through an energy landscape while in flat 
regions it is more akin to a random walk. 

\section{Results}
Having derived the breather effective potential (\ref{vpot}) we compute the 
``landscape'' corresponding to the T7 genome. 
It is natural, initially, to assume breather width is the size of a nonspecific RNAP complex. This size is not directly known, either for {\em E. coli} or
T7 RNAP. For certain other proteins the size of a nonspecific complex is 
estimated to be several times smaller \cite{IHF} than that of a specific one.
Therefore we assume upper bounds on $L$ are provided by the DNA 
``footprint'' size protected by RNAP in nuclease digestion experiments.
For translocating {\em E. coli} and T7 elongation complexes these values
are  $L_{B}=30$ bp \cite{scale1} and  $L_{\phi}=24$ bp \cite{scale2} 
respectively. 

\subsection{Sequence Analysis}
For our initial sequence analysis we adopt model parameter values coinciding 
with those of Salerno's \cite{S1} original study of T7 promoters. Setting the 
ratio $\eta = 2\times 10^-3$ implies the lower bound for breather 
width is $L_{min} \equiv \beta^{-1/2} \sim 15$ bp. 
Figure 1a shows the region of $V_{av}$  corresponding to the initial 850bp 
fragment of the T7 phage for a breather of width 30bp. For comparison Figure 
1c shows the time evolution of the system (\ref{dsc}) with breathers initially
placed at intervals within the fragment. Comparison of the three trajectories
with the effective potential landscape in Fig. 1a serves to verify that the 
direction and range of propagation agree for the two methods.\footnote{Figure 1 to go here}. 

Now the $\sigma^{70}$ subunit of {\em E. coli} holoenzyme RNAP recognises 
hexamers located 35 and 10 bases upstream of transcription initiation 
\cite{sig70}. In addition many strong promoters are enhanced \cite{UP} by a 
UP element: contacts between the $\alpha$ subunit of RNAP and AT rich 
sequences centred approximately 40-60 sites upstream. 
Inspection of Figure 1a shows that the UP region of $A_{1}$ and the -35
sites for $A_{2}$, $A_{3}$ (shown as dots) lie close to the bottom of 
potential wells: the respective initiation sites are 62, 29 and 21 bp 
upstream of these minima.  Comparison with the noise parameter, $\varepsilon$ 
plotted in Fig. 1c 
shows that when the motion is strongly deterministic ($|\varepsilon|>2$)
it is invariably towards regions where promoter recognition can occur.
The $\varepsilon$ values in the initiation region of the strongest ($A_{1}$)
bacterial promoter are 1.5 times greather than anywhere else in the T7 genome.

In fact there are seven {\em E. coli} RNAP specific promoters in the T7 
genome, the first recognition sites for the six earliest are shown in Figure 2
as dots. The four minor ($A_{0}$, $B$, $C$ 
and $E$) promoters, while having no recognised {\em in vivo} function, were 
found to have initiation sites 61 ($A_{0}$ transcribes leftwards), 27, 28 
and 17 bp downstream of deep minima.
Figure 2 also shows the full class I region of the T7 genome, transcribed by 
the bacterial RNAP, extending from the 5' DNA end to the bacterial 
transcription terminator, TE. In the Genbank \cite{GB} reference sequence 
(accession number NC\_001604) this corresponds to sites $\sim$500-7588. 
Note that other aspects of facilitated transport: dissociation 
followed by ``hopping'' or interdomain transfer are likeliest to occur in 
locally flat regions, where the breather spends most time. In this way, 
the effect of multiple, broad-bottomed wells as kinetic ``traps'' might be 
minimised. \footnote{Figure 2 to go here}

On the other hand, the deep minimum at approximately 6 Kb could be a 
desirable kinetic trap for the host RNAP as it lies 100 bp downstream of the 
gene coding for T7 RNA polymerase. The T7 RNAP intitiates transcription at
one of the two specific promoters (unfilled dots at the right side of Figure 
2) and is thus responsible for the subsequent internalisation and expression 
of the remainder of the T7 genome. This deep minimum thus represents the end 
of the region where the host RNAP is ``useful''. One finds a similar, deep 
minimum at the class II/class III interface for a wide range of breather 
widths which could play a similar role, inhibiting late transcription from 
weaker class II promoters in favour of class III promoters.

\subsection{Parameter variation} 
Given the coarseness of the current model, it is important to understand
how the results obtained may vary with respect to the parameter values. 
Up to an overall scaling, all parameter variation in the sequence-dependent 
part of (\ref{vpot}) enter via the breather width, $L$, which governs 
sensitivity to sequence-dependent inhomogeneities: an increase of width 
leads to landscapes with fewer extrema which are also broader and larger in 
amplitude. The fundamental relationship governing effects of parameter 
variations is therefore $\textrm{cosec} \mu= L \sqrt{\beta}$. It is 
natural to associate the breather family parameter $\mu$ with the dimension 
of the protein DNA-interface and $L$ with the ``response'' of the system for 
a given set of environmental conditions, encapsulated in $\eta$.

Understanding of model robustness is complicated by the 
way in which the sliding of RNAP changes. For example, if breathers do 
play a role in the location of T7 promoters  $A_{1}-A_{3}$ by bacterial RNAP 
then one might expect environmental changes which alter sliding behaviour to 
also influence promoter activity. It is known \cite{pTemp} that the 
activities of $A_{1}-A_{3}$ are temperature dependent, with $A_{1}$ 
increasing from $20-37^{\circ}$ C while initiation at $A_{2,3}$ decreases under the same circumtances 

Due to decreased thermal stability, one expects a greater ``reponse'' of the
helix to a deformation at increased temperature. For fixed $ \mu$ this 
corresponds to an increase in $L$ and decrease in $\eta$. The two graphs in 
Fig 3 are calculated for such circumstances with $L=30$, $\eta=0.002$ and 
$L=67$, $\eta=0.0004$ respectively. For the higher $\eta$ value (on the left) 
a sliding RNAP is extremely likely to fall into one of the three wells 
associated with a major promoter.\footnote{Figure 3 to go here}

Conversely for the lower $\eta$ value the well containing $A_{1}$ has greatly
widened at the expense of the other two. Indeed the -35 sites for $A_{2}$, $A_{3}$ are now situated close to local maxima and the probability of an 
encounter with sliding RNAP would be significantly reduced.
We note that minima close to one or more major promoter sites exist for a 
broad range of parameter values. One could argue that the overall 
sequence composure of the T7 initial fragment appears to confer some 
robustness of host promoter recognition against environmental variations.

Having outlined the qualitative variation of the system behaviour with 
parameter changes, we recompute the potential for $L_{\phi}=24$ bp. 
From Figure 4a it is immediately seen that for none of the T7 promoters
does the locally deepest minimum concide with upstream, recognition sites.
With replication origins $\phi_{L}$ and $\phi_{R}$ and the earliest
phage promoters, $\phi 1.1A$, $\phi 1.1B$ omitted, minima appear to be
correlated to the start of the first downstream coding sequence, as evidenced
in Fig. 4b.\footnote{Figure 4 to go here}

\section{Discussion}
\subsection{Model Assumptions}
The planar model of DNA presented is a highly simplified one, containing
numerous assumptions which are unrealistic for modelling many DNA processes:
There is no explicit allowance for the helical structure and its
writhing/twisting behaviour. Many interactions with proteins involve major,
localised conformational changes of DNA however the specific case of 
sliding RNAP may be an exception. Firstly, because such
conformational changes are unlikely to be present immediately prior to
closed complex formation \cite{y1} and secondly, there is some evidence that 
rates of RNAP sliding , under some conditions at least, are  
independent of supercoiling \cite{smeekins}.

Another important assumption was the homogeneous, harmonic nature of
the restoring torques. In fact it is known that simple, ``base content''
models of helix-coil transition thermodynamics reproduce empirical data
for short ($\leq 15$ bp) DNA oligoucleotides quite well \cite{kam2}, \cite{unif}.
Specifically, encapsulating sequence dependence as AT and GC contents 
enables reproduction of such data at 310K in 1M NaCl solution (corrections
due to change in salt concentration are discussed in \cite{unif}) with a 
mean (median) error of  9\% (5\%) (Bashford, J; unpublished). 
From our previous study of the thermodynamics of B-DNA helix-coil 
transition \cite{bas2} we further estimate that the enthalpies of A.T and
G.C pairs are in the ratio 1.56/3,  which serves to enhance the distinction 
between the two types of base pair in Eq.(\ref{vpot}). This accounts, in 
addition to differing numbers of H bonds, to the averaged effects of solution, 
neighbouring base-pairs and other interactions between the complementary pair.

The assumption of harmonicity for the stacking potential at large opening 
angles, however, is more questionable and should be further refined. 
Also molecular calculations of the ``base-flipping'' in Watson-Crick pairs 
suggest \cite{bflip} opening into the major groove is more energetically
favoured for purine bases.

\subsection{Breather dynamics}
The shape of the breather potential, used in the qualitative arguments above
depend only upon the ratios of $\eta=k/K$ and $\lambda_{A/T}/\lambda_{G/C}$. But physical properties of any breather depend on the actual parameter values. 
For example, the breather energy $E$ and oscillation frequency 
$\omega$ may be derived as
\begin{eqnarray}
E&=& \frac{16K}{L}\sim \sqrt{kK}, \\
\omega^{2} &=& \frac{K}{I}(\beta -\frac{1}{L^{2}}).
\end{eqnarray}
Using the parameter values in Ref. \cite{kl}: $K=5\times 10^{-18}$ J, 
$I=2 \times 10^{-43}$ kg m$^{2}$, in combination with our estimate
based on data from Ref.\cite{bas2}:  $k=1\times 10^{-20}$ J, yields $\eta=4.5\times 10^{-3}$.
Thus for a breather of width $L=30$ bp we get  
\begin{eqnarray*}
E \simeq 2.7 \times 10^{-18} J, & & \omega \simeq 1.0\times 10^{12} s^{-1}. 
\end{eqnarray*}
The energetic cost of creating this breather may be of the magnitude
of the electrostatic attractions responsible for the nonspecific contact.
 
Concerning the size of the DNA helix deformation, we note that parameter
$\mu$ provides an estimate of the amplitude of the base-pair opening.
$u_{max} = 4 \mu$ when $\mu <\pi/4$. For $\beta=0.0045$, as above,
the amplitude for a 30 bp breather is $2\pi/3$, corresponding to individual 
pendulum deformations of $60^{\circ}$. This parameter set does not support 
breathers of width less than $\beta^{-1/2}\simeq 15$ bp.
A variation of 20\% in the value of $K$ leads to maximum deformations of 
$52^{\circ}-65^{\circ}$: base pairs are bent but not fully opened. 
These moderate conformational changes need not be incompatible with an anticipated
absence of large deformations \cite{y1} accompanying nonspecific RNAP-DNA complexes.

The values for model parameters appearing in the literature are estimated  
from old experiments on DNA homopolymers, for example Refs. \cite{nonk}, 
\cite{yak2} which is a difficult process. However the main results of our 
paper stem from i) the {\em shape} of the potential (\ref{vpot}) and ii) the 
noise parameter, $\varepsilon$, defined by (\ref{theta}). For these two 
expressions changes in the parameter $\eta$ can be offset by ``tuning'' the 
value of $\mu$ which is a relatively free parameter. The only potentially 
serious sensitivity is that of $\varepsilon$ to large changes in $k$, the 
measure of dissociation energy for H-bonded base pairs. Fortunately, of the 
three parameters in (\ref{ifk}), this is the most reliable quantity to 
estimate.

\subsection{Helical model}
If the picture of sliding RNAP as a soliton-like deformation is subsequently 
shown to be incorrect, the correlations observed between potential minima and 
promoter sites still have to be explained. The soliton solutions of 
(\ref{ifk}) preferentially move to AT-rich regions. Inspection of (\ref{vpot})
shows the variation due to sequence is not linear in AT content, but
a first ``moment'', where the contribution from each base is weighted by its 
position relative to the central site $X$:
\begin{eqnarray}
V_{var} (X) & \sim & \sum_{i} \beta_{i} w(z_{i}),  \label{vx} \\ 
w(z)&=&\frac{\cosh z}{(\tan^{2}\mu+\cosh^{2} z)^{3/2}}. \nonumber
\end{eqnarray}
Curiously, this weighting function coincides with the inverse radius of
curvature for a hyperbolic curve $f(z)=\cosh z$. Such a term arises naturally
in the Lorenz force experienced by a charged particle following a curved magnetic field line. Initially consider a particle of mass $m$, charge 
$q$, travelling along a uniform, straight magnetic field line. Its motion is 
determined by the Lorenz equation
\begin{eqnarray*}
\frac{d}{dt}\vec{v} = \frac{q}{m} \vec{v}\times \vec{B}.
\end{eqnarray*}
Assuming the field line lies along the $z$ axis, $\vec{B}=B \vec{e}_{z}$, the
velocity equation is split into parallel and perpendicular components
\begin{eqnarray*}
\frac{d}{dt}v_{||} & = & 0, \\
\frac{d}{dt}\vec{v}_{\perp} & = & \frac{qB}{m}\vec{v}_{\perp} \times \vec{e}_{z}
\end{eqnarray*}
The general solution to these equations is a helical trajectory, with 
time-dependent coordinates
\begin{eqnarray*}
x(t) & = & x_{0}+ \frac{|v_{\perp}|}{\omega} \sin (\omega t + \phi), \\
y(t) & = & y_{0}+ \frac{|v_{\perp}|}{\omega} \cos (\omega t + \phi), \\
z(t) & = & z_{0}+ v_{||} t,
\end{eqnarray*}
where $(x_{0},y_{0},z_{0})$ denotes the initial location of the particle
and $\omega$ determines the helical frequency. 
This problem naturally resembles the electrostatic sliding of a protein 
``particle'' along the grooves of the DNA helix. Here the role of gyro 
frequency is played by the twist of the helix, while the guiding centre of 
particle motion $(x_{0},y_{0},z(t))$ corresponds to the central helical axis
of the DNA.

Consider now the effect of introducing a curve into the helical axis: a
particle travelling along a curved field line experiences a centrifugal 
force upon its guiding centre. In a local coordinate system this is
\begin{eqnarray*}
\frac{mv^{2}_{||}}{|r_{c}(s)|}\frac{\vec{r_{c}(s)}}{|r_{c}(s)|}
\end{eqnarray*}
where $|r_{c}|$ and $s$ denote the radius of curvature and line element 
along the field line.
Similarly let us here write an analogous expression
\begin{equation}
\vec{F}_{c}=\frac{{\cal E}}{r_{c}}\vec{r}_{c} 
\end{equation}
where the quantity ${\cal E}$ has the dimensions of energy. In particular,
we assume that locally the bend can be approximated by $z(\xi)=\cosh \xi$
Then, c.f. (\ref{vx}),
\begin{equation}
|\vec{F}_{c}(\xi)|={\cal E}\frac{\cosh{\xi}}{(1+\sinh^{2} \xi)^{3/2}}.
\end{equation}
It follows that in the continuum limit the time-averaged breather potential 
could also be thought of as the work done by a ``centrifugal force'' on a
sliding RNAP as it navigates a bend in the helix.
Therefore the ``potential'' (\ref{vpot}) can conceivably be arrived at
via simple considerations of thermal stability (in a planar model) or bending 
deformations (in a helical model), two of the most commonly suggested
 mechanisms for enhancing promoter recognition.

\subsection{Superhelicity}
A mechanism of localised DNA deformation with demonstrated biological 
significance \cite{ben3}, \cite{ben1},\cite{ben2} is that of superhelical 
stress-induced DNA denaturation (SSID). Roles for SSID in gene
regulation have been proposed \cite{ben3} in regard to both open complex 
formation and transcription. In the former instance, promoter sites are 
easily destabilized by superhelical stress. In the latter, 
the action of local helix unwinding by transcribing RNAP results in waves of 
positive (negative) superhelicity propagating downstream (upstream) of the 
transcription complex. Computation of SSID profiles indicates \cite{ben3},
\cite{ben2} AT rich regions (down-) up-stream of the (3') 5' ends of 
transcription units are prone to localised over/under-winding  acting as a 
possible ``sink'' for propagating superhelicity and ensuring smooth transcription.

The breather potential (\ref{vpot}), which also picks out regions of AT 
shows that transcription units of at least $10^{3}$ bp in length are often 
demarcated by minima, in agreement with the above observations.
This is especially the case for the 3' ends of T7 genes 1 and 6, the last
genes in class I and II regions respectively. In these instances the AT 
richness may also confer extra rigidity, making these suitable pause sites 
in the stepwise internalisation of the phage genome, or as mentioned above
act as a kinetic trap, used in inhibiting class I or II transcription.

\subsection{Correlations}
In reporting promoter-extrema correlations two points should be kept in mind.
Firstly, the assumed breather widths coincide with the sizes of the elongation
RNAP-DNA complexes. Therefore potential minima could be indicative of 
deformation associated with transcription, as appears to be the case for T7
phage promoters, shown in Figure 4. Regarding nonspecific complexes, the 
values $L_{B}=30$ and $L_{\phi}=24$ bp should be considered as upper bounds 
for an experimentally undetermined quantity.
The correlations reported in this study persist for the ranges 
$20\leq L_{B}\leq 30$ and $18\leq L_{\phi}\leq 24$. For sizes less than 18bp, 
the increasing roughness of Eq.(\ref{vpot}) causes difficulty in identifying 
correlations.

The second caveat is that only correlations between promoter initiation 
and the deepest local minimum have been considered. For some T7 promoters 
shallow upstream wells also exist. Moreover the effect of thermal noise has 
not been considered. Only with full dynamical simulations can connections
between the local topography of Eq.(\ref{vpot}) and facilitated target 
location be properly studied.

It is difficult to see how kink solutions of the planar model 
(\ref{ifk}), previously considered \cite{S1}-\cite{sanchez} might mimic 
physical profiles of base-pair opening. Kinks will also move preferentially to AT rich regions, presumably the reason why promoter sequences $A_{1}$ \cite{S1}, $A_{3}$ and $A_{0}$ \cite{S2} were concluded to be ``dynamically active''.
The unit-mass potential for kinks, initially at rest, moving in a 
slowly-varying background was derived by Salerno and Kivshar \cite{S3}. The sequence variation is contained in a term analogous to (\ref{vx}), however the 
weighting function is 
\begin{eqnarray*}
W_{k}(z)=\textrm{sech}^2 z.
\end{eqnarray*}
This coincides with the breather function for small $\tan^{2} \mu$, 
illustrating why similar results for the major T7 promoter sequences
are obtained for both kink \cite{S1}-\cite{S3} and breather solitons.

\section{Conclusion}
In this paper we have re-examined Salerno's nonlinear DNA model, postulating 
a role for localised soliton excitations in approximating the sliding 
component of facilitated target location of RNA polymerase.
We found that such deformations would involve moderate bending of individual 
base pairs and that their energy of translocation is consistent with a picture 
of noisy, deterministic dynamics. Both of these observations are also 
consistent with current, limited knowledge of RNAP sliding and nonspecific 
complexes. A qualitative correspondence of these solitons and localised 
bending in a helical model was also demonstrated.

The dynamical picture of sliding which emerged also suggests that the 
random/deterministic nature of the motion is sequence-dependent, with 
translocation in relatively homogeneous regions being effectively random.
The corollary, that interplay between adjacent random and deterministic 
regions could constitute a search ``algorithm'', is speculative and, we 
believe, merits further investigation.

Our analysis of the T7 genome showed good correlations between AT-rich 
regions and the recognition sites of host-specific promoters used for 
early phage transcription. For phage-specific promoters, regions of 
maximal AT-richness correlated with the start of the coding sequence 
immediately downstream. As discussed above this may be connected with 
transcription and while there is no obvious correlation with recognition 
sites, a full description of facilitated target location needs to account
for the thermal background. This is a subject of current investigation.

We note that there has been suggestion \cite{mol4} that virion proteins 
injected into the host cell with the initial T7 fragment may i) inhibit the 
nonspecific binding of restriction enzymes and other proteins to DNA;  ii) 
have an affinity for {\em E. coli} RNAP, negating the requirement for direct 
promoter recognition {\em in vivo}. Similarly, inhibition of class I and II 
transcription is known to be performed by T7 gene products: kinase (gene 0.7) 
and lysozyme (gene 3.5) respectively.

However we see similar correlations for the UP and $\sigma^{70}$ sites of 
bacterial promoters in other members of the T7 viral supergroup, in 
addition to genomes of the unrelated phages T4 and T5 (see Figure 5).
This may be suggestive of a mechanism at work to enhance promoter 
recognition/inhibition in lytic phage genomes, although in the presence of 
functional proteins this mechanism can be relegated to an auxiliary role, 
such as in T7. \footnote{Figure 5 to go here}

It is important to investigate whether planar base-flipping/helical bending 
deformation patterns can be used to simulate protein-DNA interactions in 
DNA sequence analysis. The correlations reported here, to our knowledge for 
the first time, could have been made via other ``nonlinear'' analyses of AT 
content, had a motivation been apparent. 
Propagation of breathers in a non-linear, toy model of DNA provide a source, 
for such motivation. It may be that herein lies the true value of a model 
with such a controversial history.

\vspace{0.5cm}
\noindent
\large{ {\bf Acknowledgements}}

\noindent
This research was funded by Australian Research Council grant DP0344996 and a 
visiting fellowship to the Centre for Nonlinear Physics, Australian National 
University, where part of this work took place. 
 The author thanks G. Yang for helpful remarks and is grateful to Yu. Kivshar
and I. Molineux for discussions and comments on earlier versions of the manuscript.

\section{Figure Captions}
\noindent
{\bf Figure 1:}
a) Effective potential (\ref{vpot}) for breathers in the initial T7 virion 
fragment. Initial binding sites for bacterial promoters are denoted by
dots; b) Noise parameter $\varepsilon(X)$ for the same sequence.
c) Evolution over 1000 time-steps of the system (\ref{dsc}) with 
breathers initially placed at sites 460, 570 and 680.
\\

\noindent
{\bf Figure 2:}
Effective potential (\ref{vpot}) for 30bp wide breathers in the class I region of the T7 genome. Filled and unfilled dots denote respectively UP or -35 {\em E. coli} and +1 T7 promoter sites. \\

\noindent
{\bf Figure 3:}
Potential (\ref{vpot}) computed for the T7 initial fragment
for $\mu=\pi/6.05$.
a) $\eta=0.002$, $L=30$ bp; b) $\eta=0.0004$ ($L=67$ bp); 
Dots denote, from left to right, UP and -35 sites for $A_{1}-A_{3}$ 
bacterial promoters.\\

\noindent
{\bf Figure 4:}
a) Location of minima of (\ref{vpot}) nearest initiation sites of T7 phage 
promoters; b) Scatter plot of initiation-downstream transcription unit 
distance (TU) versus initiation minima distance (Min).\\

\noindent
{\bf Figure 5:}
Representative region of T5 genome potential, showing correlations between 
potential minima and -35 sites for {\em E. coli} promoters ($L=30$ bp).

\begin{figure}[tbp]
 \centering{
\resizebox{14cm}{8cm}{\includegraphics{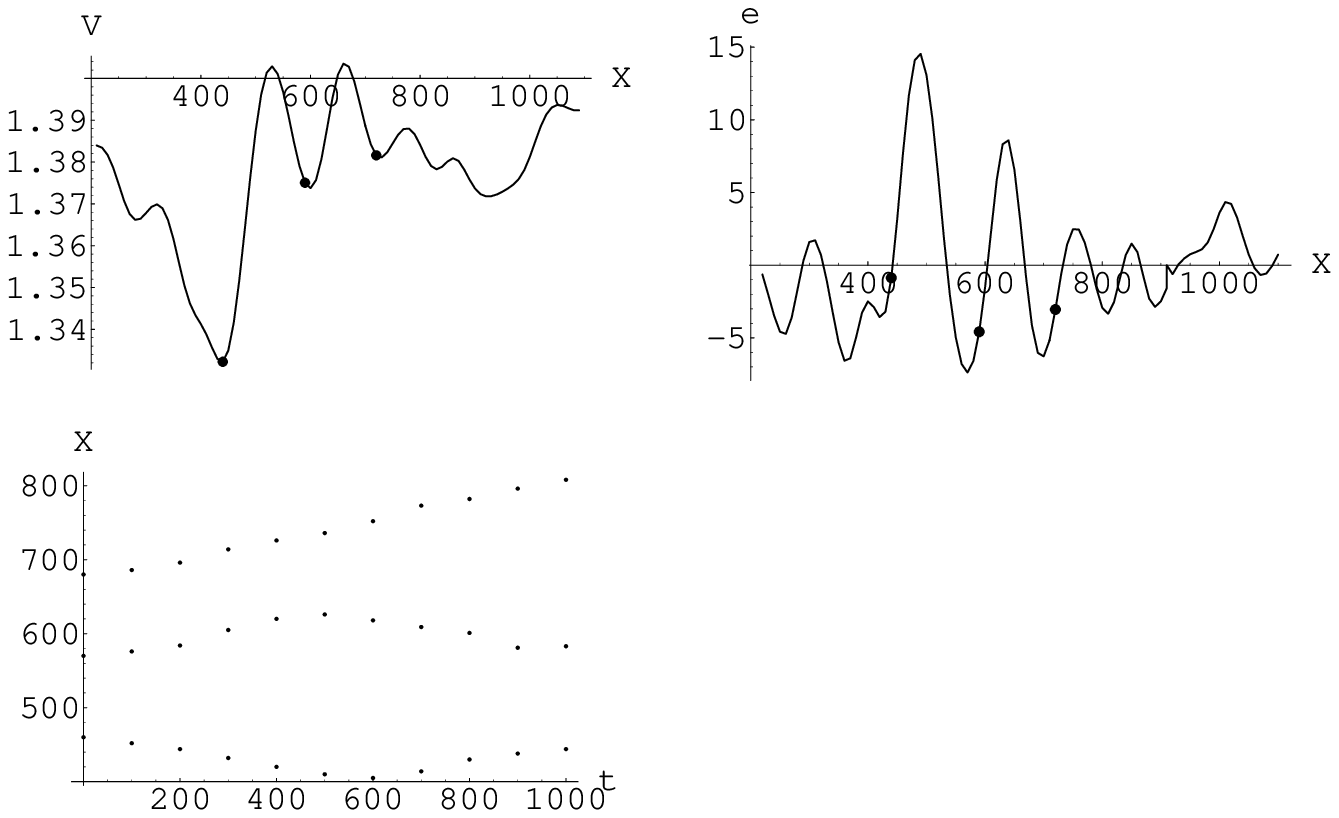}}
}
\caption{}
\protect \label{j1}
\end{figure}

\begin{figure}[htb]
 \centering{
\resizebox{11cm}{6cm}{\includegraphics{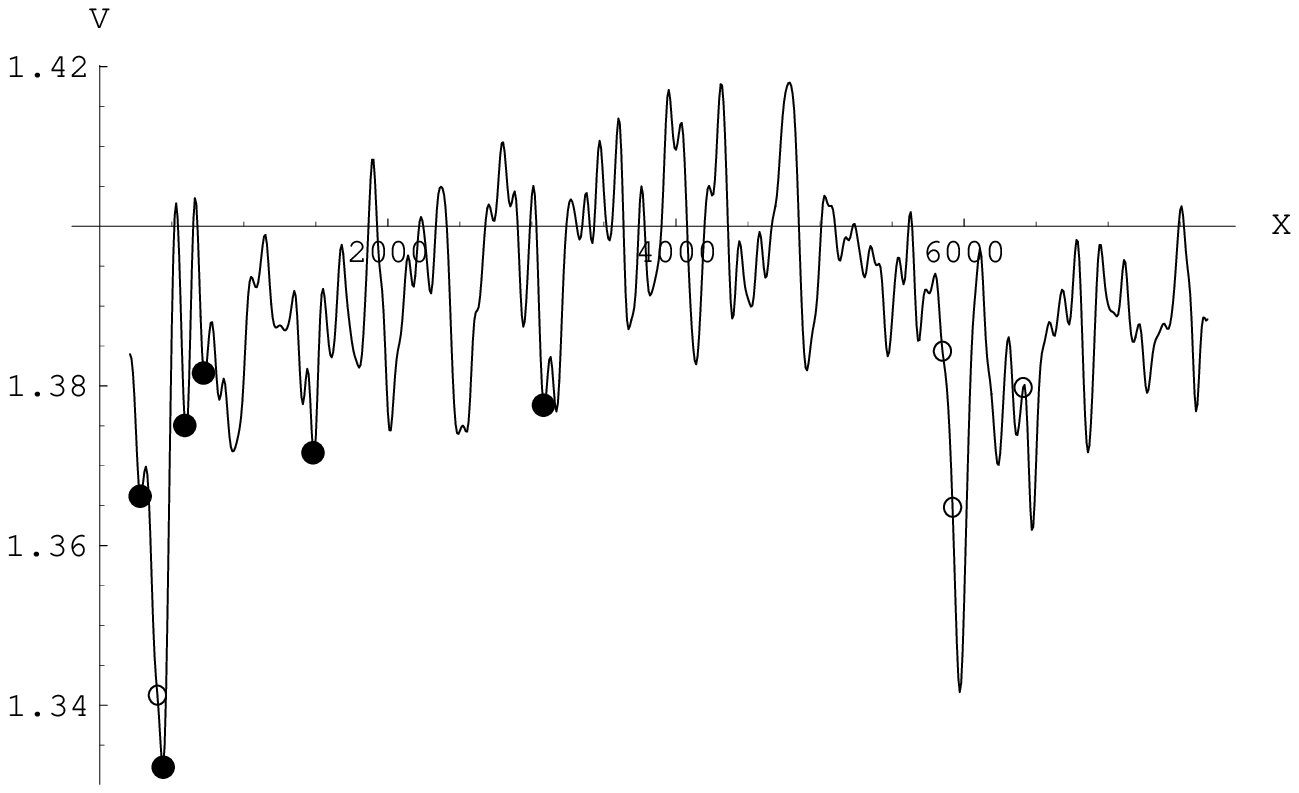}}
}
\caption{}
\protect \label{j2}
\end{figure}

\begin{figure}[htb]
 \centering{
\resizebox{14cm}{4cm}{\includegraphics{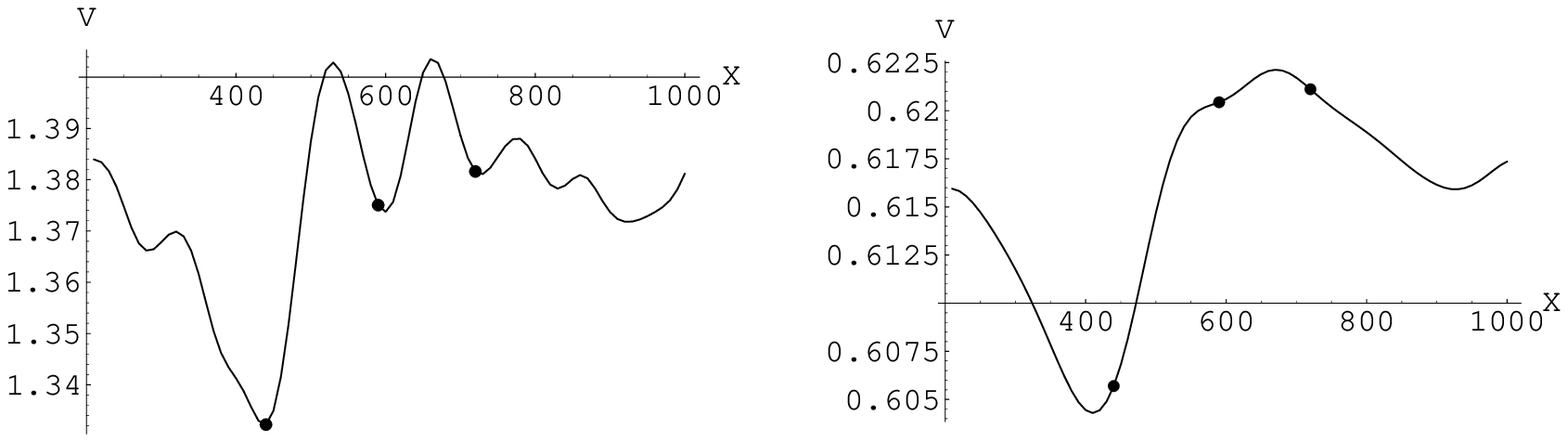}}
}
\caption{}
\protect \label{j3}
\end{figure}

\begin{figure}[htb]
 \centering{
\resizebox{14cm}{4cm}{\includegraphics{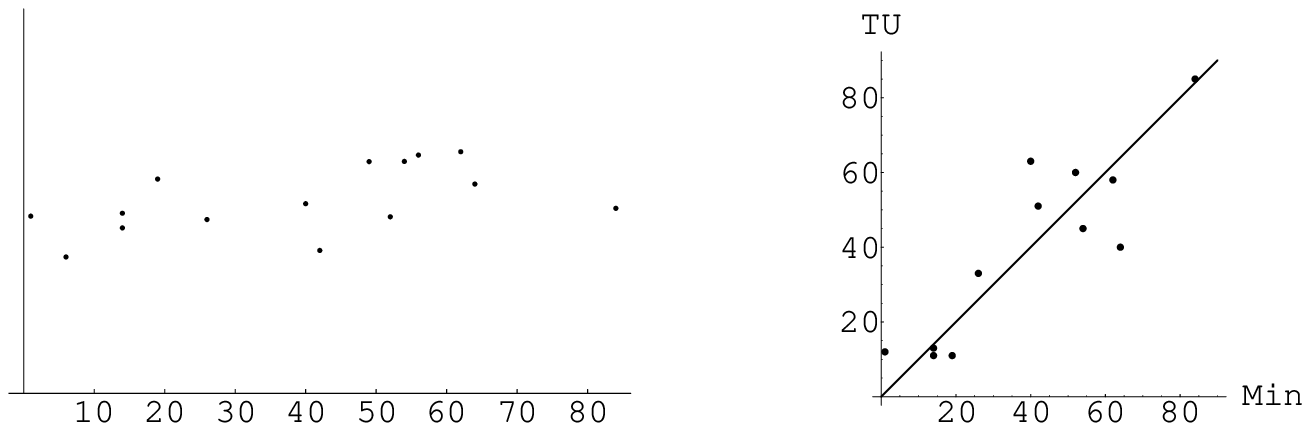}}
}
\caption{}
\protect \label{j4}
\end{figure}

\begin{figure}[htb]
 \centering{
\resizebox{12cm}{7cm}{\includegraphics{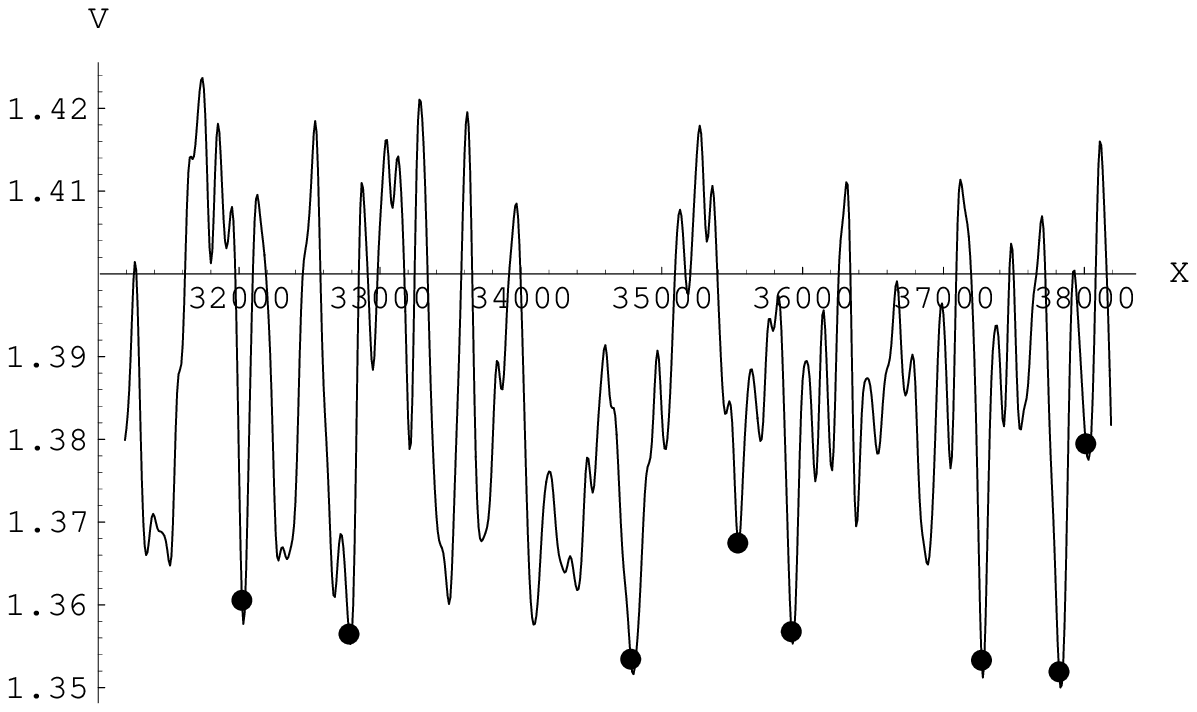}}
}
 \caption{}
\protect \label{j5}
\end{figure}


\begin{thebibliography}{99}
\bibitem{pey}  Peyrard, M. ``Nonlinear dynamics and statistical physics of DNA'', {\em Nonlinearity} {\bf 17} (2004), R1-R40.
\bibitem{Gaeta1} Gaeta, G. 
``Results and limitations of the soliton theory of DNA transcription'', {\em J. Biol. Phys.} {\bf 24} (1999), 81-96.
\bibitem{PB1} Peyrard, M. and Bishop, A.R. 
``Statistical mechanics of a nonlinear model for DNA denaturation'', {\em Phys. Rev. Lett.} {\bf 62} (1989), 2755-2758.
\bibitem{Yak} Yakushevich, L.V. ``Is DNA a nonlinear dynamical system where solitary conformational waves are possible?'', {\em J. Biosci.} {\bf 26} (2001), 305-313.
\bibitem{dinuc} Bruant, N., Flatters, D., Lavery, R. and Genest, D. 
``From atomic to mesoscopic descriptions of the internal dynamics of DNA'', {\em Biophys. J.} {\bf 77} (1999), 2366-2376. 
\bibitem{S1} Salerno, M. ``Discrete model for DNA-promoter dynamics'', {\em Phys. Rev.} {\bf A44} (1991), 5292-5297.
\bibitem{S2} Salerno, M. ``Dynamical properties of DNA promoters'', {\em Phys. Lett.} {\bf A167} (1992), 49-53.
\bibitem{S3} Salerno, M. and Kivshar, Yu.S. ``DNA promoters and nonlinear dynamics'', {\em Phys. Lett.} {\bf A193} (1994), 263-266.
\bibitem{S3b} Salerno, M. ``Nonlinear dynamics of plasmid PBR322 promoter'',
chapter 10 in  M. Peyrard (ed.), {\em Nonlinear excitations in biomolecules}, Edition de Physique, Springer, New York (1995).
\bibitem{kl} Lennholm and E.; H\"{o}rnquist, M. ``Revisiting Salerno's sine-Gordon model of DNA: active regions and robustness'', {\em Physica} {\bf D177} (2003), 233-241.
\bibitem{sanchez} Cuenda, S., S\'{a}nchez, A. 
``Disorder and fluctuations in nonlinear excitations in DNA'', {\em Fluct. Noise Lett.} {\bf 4} (2004), L491-L504. 
\bibitem{englander}
Englander, S.W. {\em et al.} ``Nature of the open state in long polynucleotide double helices: possibility of solition excitations'', {\em Proc. Natl. Acad. Sci.} {\bf 77} (1980), 7222-7226.
\bibitem{riv} Gaeta. G., Reiss, C., Peyrard, M. and Dauxios, T. ``Simple models of nonlinear DNA dynamics'', {\em Riv. del. Nuov. Cim.} {\bf 17} (1994), 1-48.
\bibitem{edwards} Gabriel, C. {\em et al.} ``Microwave absorption in aqueous solutions of DNA'', {\em Nature} {\bf 328} (1987) 145-146. 
\bibitem{bigio} Bigio, I.J., Gosnell, T.R., Mukherjee, P. and Safer, J.D. 
``Microwave absorption spectroscopy of DNA'', {\em Biopolymers} {\bf 33} (1993), 147-150. 
\bibitem{gueron} Gu\'{e}ron, M., Kochoyan, M. and Leroy, J.L. ``A single mode of DNA base-pair opening drives imino proton exhange'', {\em Nature} {\bf 328} (1987), 89-92.
\bibitem{Kam} Frank-Kamensteskii, M. ``Physicists retreat again'', {\em Nature}  {\bf328} (1987), 108.
\bibitem{ben3} Benham, C.J. ``Duplex destabilization in superhelical DNA is 
predicted to occur at specific transcriptional regulatory regions'', {\em J. Mol. Biol.} {\bf255} (1996), 425-434.
\bibitem{PB2} Choi, C.H. {\em et al.} ``DNA dynamically directs its own transcription initiation'', {\em Nucl. Acids. Res.} {\bf 32} (2004), 1584-1590.
\bibitem{bmc} Kanhere, A. and Bansal K. ``A novel method for prokaryotic promoter prediction based on DNA stability'', {\em BMC Bioinformatics} {\bf 6} (2005), 1-10.
\bibitem{trif} Bolshoy, A., McNamara, P., Harrington, R.E. and Trifonov, E. ``Curved DNA without A-A: experimental estimation of all 16 DNA wedge angles'', {\em Proc. Natl. Acad. Sci.} {\bf 88}, (1991) 2312-2316.
\bibitem{anselmi} Scipioni, A. {\em et al.}
``Sequence-dependent DNA curvature and flexibility from scanning force microscopy images'', {\em Biophys. J.} {\bf 83} (2002), 2408-2418.
\bibitem{lankas} Lankas, F. ``DNA sequence-dependent deformability - insights from computer simulations'', {\em Biopolymers} {\bf 73} (2004), 327-339.
\bibitem{naplus} Ponomarev, S.Y., Thayer, K.M., Beveridge, D.L. ``Ion motions in molecular dynamics simulations on DNA'', {\em Proc. Natl. Acad. Sci.} {\bf 101} (2005), 14771-14775.
\bibitem{proz} Polozov, R.V. {\em et al.} ``Electrostatic potentials of DNA. Comparative analysis of promoter and nonpromoter sequences.'', {\em J. Biomol. Struct. Dyn.} {\bf 16} (1999), 1135-1143.
\bibitem{kbk} Braun, O.M. and Kivshar, Yu.S.: {\em The Frenkel-Kontorova Model: Concepts, Methods and Applications}, Springer, Berlin, 2004.
\bibitem{vonhip} von Hippel, P.H. and Berg, O.G. ``Facilitated target location in biological systems'', {\em J. Biol. Chem.} {\bf 264} (1989), 675-678.
\bibitem{lac1} Fickert, R. and M\"{u}llerhill, B. ``How lac repressor finds {\em lac} operator {\em in vivo}'', {\em J. Mol. Biol.} {\bf 226} (1992), 59-68.
\bibitem{p53} Jiao, Y., Cherny, D.I.,  Heim, G., Jovin, T.M. and Sch\"{a}ffer, T.E.`` Dynamic interactions of p53 with DNA in solution by time-lapse atomic force microscopy'', {\em J. Mol. Biol.} {\bf 314} (2001), 233-243.
\bibitem{wu} Park, C.S., Wu, F.Y.H. and Wu, C.S. ``Molecular mechanism of 
promoter selection in gene transcription'', {\em J. Biol. Chem.} {\bf 257} (1982), 6950-6956.
\bibitem{wu2} Singer, P.T. and Wu, C.S. ``Kinetics of promoter search by {\em Escherichia coli} RNA polymerase'' {\em J. Biol. Chem.} {\bf 263} (1988), 4208-4214.
\bibitem{smeekins} Smeekins, S.P. and Romano, L.J. ``Promoter and nonspecific DNA binding by the T7 RNA polymerase'', {\em Nucl. Acids. Res.} {\bf 14} (1986), 2811-2827.
\bibitem{kabata} Kabata, H. {\em et al.} ``Visualisation of single molecules of RNA polymerase sliding along DNA'', {\em Science} {\bf 262} (1993), 1561-1563.
\bibitem{bustamente} Guthold, M. {\em et al.} 
``Direct observation of one-dimensional diffusion and transcription by {\em Escherichia coli} RNA polymerase'', {\em Biophys. J.} {\bf 77} (1999), 2284-2294.
\bibitem{ecorv} Jeltsch, A. and Pingoud, A. 
``Kinetic characterisation of linear diffusion of the restriction endonuclease
{\em Eco}RV on DNA'', {\em Biochemistry} {\bf 97} (1998), 2160-2169.
\bibitem{meth} Nardone, G., George, J. and Chirikjian, J.G. ``Differences in the kinetic properties of BamH1 endonuclease and methylase with linear DNA substrates'', {\em J. Biol. Chem.} {\bf 261} (1986) 2128-2133.
\bibitem{berg} Berg, O.G., Winter, R.B. and von Hippel, P.H. 
``Diffusion-driven mechanisms of protein translocation on nucleic acids. 1. Models and Theory'', {\em Biochemistry} {\bf 20} (1981), 6929-6948.
\bibitem{S4} Barbi, M., Place, C., Popkov, V. and Salerno, M. 
``A model of sequence-dependent protein diffusion along DNA'', {\em J. Biol. Phys.} {\bf 30} (2004), 203-226.
\bibitem{satar} Satari\`{c}, M.V. and Tuszy\`{n}ski, J.A. ``Impact of regulatory proteins on the nonlinear dynamics of DNA'', {\em Phys. Rev.} {\bf E65} (2002), 1901-1911.
\bibitem{ting0} Ting, J.J-L. and Peyrard, M. ``Effective breather-trapping mechanism for DNA transcription'' {\em Phys. Rev.} {\bf E53} (1996), 1011-1018.
\bibitem{ting1} Ting, J.J-L. ``DNA transcription mechanism with a moving enzyme'', {\em Intl. J. Mod. Phys.} {\bf A7} (1997), 1125-1132.
\bibitem{endy} Endy, D., You, L., Yin J. and Molineux, I.J. ``Computation, predictions and experimental tests of fitness for bacteriophage T7 mutants with permuted genomes'', {\em Proc. Natl. Acad. Sci.} {\bf 97} (2000), 5375-5380. 
\bibitem{hes} Hesselbach, B.A. and Nakada, D. ```Host shut off' function of bacteriophage T7: involvement of T7 gene 2 and gene 0.7 in the inactivation of {\em Escherichia coli} RNA polymerase'',{\em J. Virol} {\bf 24} (1977), 736-745.
\bibitem{lys} Moffat, B.A. and Studier, F.W. ``T7 lysozyme inhibits transcription by T7 RNA polymerase'', {\em Cell} {\bf 49} (1987), 221-227.
\bibitem{inf1} Zavriev, S.K. and Shemyakin. M.F.
``RNA polymerase-dependent mechanism for the stepwise T7 phage DNA transport from the virion into {\em E. coli}'', {\em Nucl. Acids. Res.} {\bf 10} (1982), 1635-1652.
\bibitem{inf2} Garcia, L.R., and Molineux, I.J. ``Rate of translocation of bacteriophage T7 DNA across the membranes of {\em Escherichia coli}'', {\em J. Bacteriol.} {\bf 177} (1995), 4066-4076.
\bibitem{zhang} Zhang, F. ``Breather scattering by impurities in the sine-Gordon model'', {\em Phys. Rev.} {\bf E58} (1998), 2558-2563.
\bibitem{IHF} Tsodikov, O.V., Holbrook, J.A., Shkel, I.A., and Record, M.T., Jnr. ``Analytic binding isotherms describing competitive interactions of a protein ligand with specific and nonspecific sites on the same DNA oligomer'',
 {\em Biophys. J.} {\bf 81} (2001), 1960-1969.
\bibitem{scale1} von Hippel, P.H. ``An integrated model of the transcription complex in elongation, termination and editing'', {\em Science} {\bf 281} (1998), 660-665.
\bibitem{scale2} Imburgio, D., Rong, K. Ma. and McAllister, W.T. ``Studies of promoter recognition and start site selection by T7 RNA polymerase using a comprehensive collection of promoter variants'', {\em Biochemistry} {\bf 39} (2000), 10419-10430.
\bibitem{sig70}  Mulligan, M.E., Hawley, D.K., Entriken, R. and McClure, W.R. 
``{\em Escherichia coli} promoter sequences predict {\em in vitro} RNA polymerase selectivity'', {\em Nucleic. Acids. Res.} {\bf 12} (1984), 789-800.
\bibitem{UP} Estrem, S.T. {\em et al.} ``Bacterial promoter architecture: subsite structure of UP elements and interactions with the carboxy-terminal domain of the RNA polymerase $\alpha$ subunit'', {\em Genes Dev.} {\bf 13} (1999), 2134-2147.
\bibitem{A1} Sclavi, B. {\em et al.} ``Real-time characterisation of intermediates in the pathway to open complex fomration by {\em Escherichia coli} RNA polymerase at the T7A1 promoter'', {\em Proc. Natl. Acad. Sci.} {\bf 102} (2005), 4706-4711.
\bibitem{GB} National Center for Biotecnhnology Information website. http://www.ncbi.nlm.nih.gov/Entrez
\bibitem{pTemp} Dausse, J.P., Sentenac, A. and Fromageot, P. 
``Interaction of RNA polymerase from {\em Escherichia coli} with DNA. Effect of temperature and ionic strength on selection of T7 DNA early promoters.''
{\em Eur. J. Biochem} {\bf 65} (1976), 387-393.
\bibitem{bflip} Giudice, E., V\'{a}rnai, P. and Lavery, R. 
``Base-pair opening within B-DNA: free energy pathways for GC and AT pairs from umbrella sampling situations'', {\em Nucl. Acids. Res.} {\bf 31} (2003), 1434-1443.
\bibitem{y1} Murakami, K.S., Masuda, S. and Darst, S.A. ``Structural basis of transcription initiation: RNA polymerase holoenzyme at 4 \AA resolution'', {\em Science} {\bf 296} (2002), 1280-1284.
\bibitem{kam2} Frank-Kamanetskii, M. ``Simplification of the empirical relationship between melting DNA, its GC content and concentration of sodium ions in solution'', {\em Biopolymers} {\bf 10} (1971), 2623-2624.
\bibitem{unif} SantaLucia, J. Jnr. 
``A unified view of polymer, dumbbell and oligonucleotide DNA nearest-neighbour thermodynamics'', {\em Proc. Natl. Acad. Sci.} {\bf 95} (1998), 1460-1465.
\bibitem{bas2} Bashford, J.D. and Jarvis, P.D. ``A base-pairing model of duplex formation I: Watson-Crick pairing geometries'', {\em Biopolymers} {\bf 78} (2005), 287-297.
\bibitem{santa} SantaLucia, J. Jnr., Allawi, H.T. and Seneviratne, P.A. ``Improved nearest-neighbour parameters for predicting DNA duplex stability'', {\em Biochemistry} {\bf 35} (1996), 3555-3562.
\bibitem{nonk} Yakushevich, L.V. ``Scattering of neutrons and light by DNA solitons'', {\em Stud. Biophys.} {\bf 103} (1984), 171-178.
\bibitem{yak2} Yakushevich, L.V. ``The effects of damping, external fields and inhomogeneity on the nonlinear dynamics of bioploymers'', {\em Stud. Biophys.} {\bf 121} (1987), 201-207.
\bibitem{mol4} Molineux, I.J. ``No syringes please, ejection of phage T7 DNA from the virion is enzyme driven'', {\em Mol. Microbiol.} {\bf 40} (2001), 1-8.
\bibitem{ben1} Benham, C.J. ``Sites of predicted stress-induced DNA duplex destabilization occur preferentially at regulatory regions'', {\em Proc. Natl. Acad. Sci.} {\bf 90} (1993), 2999-3003.
\bibitem{ben2} Wang, H., Noordewier, M. and Benham, C.J. ``Stress-Induced DNA Duplex Destabilization (SIDD) in the {\em E. coli} genome: SIDD sites are closely associated with promoters'', {\em Genome Research} {\bf 14} (2004), 1575-1584.

\end{thebibliography}
\end{document}